\documentstyle[12pt]{article}
\date{}
\def\beq{\begin{equation}}
\def\eeq{\end{equation}}
\def\beqa{\begin{eqnarray}}
\def\eeqa{\end{eqnarray}}

\begin{document}
\title{The Anharmonic Correction in the Soliton Model of the Hyperons}

\author{M. Bj\"{o}rnberg, K. Dannbom and D.O. Riska}
\maketitle

\centerline{\it Department of Physics, P.O. Box 9, 00014 University of
Helsinki, Finland}

\setcounter{page}{0}
\vspace{1cm}

\centerline{\bf Abstract}
\vspace{0.5cm}

We derive the anharmonic correction to the hyperon energy in the bound
state version of the topological soliton model for the hyperons, and
show that it represents a negative correction of at most 10\% to the
energy of the bound heavy flavour two-meson system in the case of cascade
hyperons. The main anharmonic correction arises from the mass term
in the Lagrangian density. For large meson masses the consistency of the
model requires that
the anharmonic correction decreases as the inverse
square root of the
mass of the heavy flavour meson.

\newpage

\centerline{\bf 1. Introduction}
\vspace{0.5cm}

The bound state version of the topological soliton model for the
hyperons originally proposed by Callan and Klebanov [1] has been shown
to provide a remarkably good description of the static properties of
the hyperons in all heavy flavour generations [2-4]. The version of
this model in which the soliton is stabilized by explicit vector meson
fields [3] has been shown to provide a smooth interpolating model, which
possesses chiral symmetry in the light flavour sectors and heavy quark
symmetry [6] in the limit when the heavy meson mass grows beyond
limit [7,8]. As the model thus has all the symmetries that are
believed to be important in the low energy sector, it should be viewed
as realistic enough to have predictive power - as e.g. in its
prediction of the existence of nonstrange pentaquark states [9].\\

The original version of the bound state soliton model [1] is based on
the direct extension of Skyrme's Lagrangian density [10] to $SU(3)$,
and an expansion to second order of the isodoublet kaon field which is
treated as a quantum fluctuation. The
resulting harmonic approximation leads to a wave equation for the kaon
field, which has two bound states, the lower one of which describes the
$\Lambda-\Sigma$ system ($P$-state) and the excited one the
$\Lambda(1405)$
($S$-state) [1,2,11].
The cascade and $\Omega^{-}$ particles are then constructed as
multimeson states, which in the harmonic approximation do not
interact. In this paper we investigate the validity of this harmonic
approximation by carrying out the expansion in the meson field to
quartic order. Because of the complicated form of the stabilizing term
in the Lagrangian density of the Skyrme model [10] we employ the
simpler alternate form proposed by Pari [12]. This alternative
stabilizing term, which is quartic in the derivatives, is equivalent
to Skyrme's quartic stabilizing term at the level of $SU(2)$, but not
in the case of $SU(3)$ fields. In the case of the kaons it has been
shown that linear combinations of the two alternate quartic stabilizing
terms lead to good predictions for the strange hyperon energies,
although if the alternate (Pari) term is employed alone
bound $\eta N$ states would appear, in contradiction with
empirical findings [13]. As the kaon bound state energy is fairly
insensitive to the choice of quartic stabilizing term we shall however here be
content to consider only the simpler alternate quartic stabilizing
term, as this should suffice for the purpose of estimating the
anharmonic correction.\\

The anharmonic term of lowest order takes the form of a meson-meson
interaction, which we treat in lowest order perturbation theory. The
matrix elements of this interaction for two-meson hyperon states - i.e. the
cascade particles and their heavy flavour analogs - represent the
perturbative estimate
of the anharmonic correction to the hyperon mass values, the quality of
which should be comparable to the standard perturbation theory
estimate of the electron-electron interaction energy in the case of
the $He$-atom. We find that in the case of the cascade type hyperons the
anharmonic correction amounts to a negative correction of at most 10\%
to the energy of the bound two meson system. In the case of the strange cascade
hyperons the correction is -26 MeV, in the case of the $\Xi_c$ -233
MeV and in the case of the predicted $\Xi_b$ hyperons -639 MeV when
standard values for the parameters of the model are used. We
show that in the case of very large meson mass values
the consistency of the model requires that the anharmonic
correction is roughly proportional to the
inverse square root of the heavy meson mass. The relative smallness of
the anharmonic correction indicates that the harmonic approximation is
reliable and that the satisfactory predictions that it gives for the
hyperon spectra and magnetic moments are fairly robust.\\

This paper is organized into sections. In section 2 we derive the
general form for the lowest anharmonic term in the Lagrangian density
of the meson-soliton system. In section 3 we show that the matrix
element of this term in the case of a state with two mesons in the
ground state can be
reduced to a fairly simple form. In section 4 we derive an approximate
estimate for the dependence on the heavy flavour meson mass of the
anharmonic correction. In section 5 we present the numerical results
for the anharmonic correction to the cascade-type hyperons. Section 6
contains a summarizing discussion.\\

\vspace{1cm}

\centerline{\bf 2. The anharmonic Hamiltonian density}
\vspace{0.5cm}

{\bf 2.1 The harmonic Lagrangian}
\vspace{0.5cm}

In its simplest version the bound state model for the hyperons is
based on a Lagrangian density for an $SU(3)$-valued field $U$ that has
three components:

$${\cal L}={\cal L}_\sigma+{\cal L}_S+{\cal L}_{CSB}.\eqno(2.1)$$
The first of these is the Lagrangian density of the
non-linear $\sigma$-model:

$${\cal L}_\sigma=-{1\over 4}f_\pi^2Tr L_\mu L^\mu,\eqno(2.2)$$
where the "left current" $L^\mu$ is defined as

$$L^\mu=U^\dagger \partial^\mu U.\eqno(2.3)$$
The second term ${\cal L}_S$ is a stabilizing term for which a variety
of forms has been considered [10,12,14]. We shall here employ the form

$${\cal L}_{S}={1\over 16 e^2}\{(Tr L_\mu L_\nu)^2-(Tr L_\mu
L^\mu)^2\},\eqno(2.4)$$
that was suggested by Pari [12] as a simpler, although $SU(2)$
equivalent, alternative to Skyrme's original quartic stabilizing term
[10]. The third term ${\cal L}_{CSB}$ takes into account the chiral
symmetry breaking meson mass term and the $SU(3)$ symmetry breaking
implied by the different values for the decay constants of the mesons
in the different flavour generations. For this term we employ the form
[15]

$${\cal
L}_{CSB}=\frac{f_\pi^2m_\pi^2+2f^2m^2}{12}Tr\{U+U^\dagger-2\}$$
$$+\frac{f_\pi^2m_\pi^2-f^2m^2}{6}Tr\{\sqrt{3}\lambda_8(U+U^\dagger)\}$$

$$-\frac{f_\pi^2-f^2}{12}Tr\{(1-\sqrt{3}\lambda_8)(U\partial_\mu U^\dagger
 \partial^\mu U+U^\dagger \partial_\mu U \partial^\mu
U^\dagger)\},\eqno(2.5)$$
where $m$ represents the mass of the heavy flavour meson $(K,\, D$ or
$B$) and $f$ the corresponding decay constant.\\

This Lagrangian density has to be augmented by the Wess-Zumino action

$$S_{WZ}=-{iN_C\over 240\pi^2}\int d^5
x\epsilon^{\mu\nu\alpha\beta\gamma}Tr\{L_\mu L_\nu L_\alpha L_\beta
L_\gamma\},\eqno(2.6)$$
which contributes to the energy when the field manifold is larger than
$SU(2)$, although it cannot be reduced to the form of a Lagrangian
density [1,2]. Here $N_C$ is the colour number.\\

For the $SU(3)$ field $U$ we adopt the form [1]

$$U=\sqrt{U_\pi}U_m \sqrt{U_\pi},\eqno(2.7)$$
where $U_\pi$ is the soliton field

$$U_\pi=\left( \begin{array}{cc} u & 0\\ 0 & 1 \end{array}
\right).\eqno(2.8)$$
In this expression $u$ is the usual $SU(2)$ hedgehog field:

$$u=e^{i\vec \tau \cdot \hat r \theta(r)},\eqno(2.9)$$
where $\theta$ is the chiral angle.\\

The heavy meson field $U_m$ has the form

$$U_m=exp\{i{\sqrt{2} \over f}\left(\begin{array}{cc} 0 & M\\
M^\dagger & 0 \end{array} \right) \}.\eqno(2.10)$$
Here $M$ represents one of the meson isodoublets $(K^+,K^0)^T$, $(\bar
D^0,D^{-})^T $ or $(B^+,B^0)^T$ [4].\\

The bound state model for the hyperons is obtained by expanding the
Lagrangian density (2.1) to second order in the meson field $M$. This
harmonic approximation leads to a linear differential equation of
second order for the meson field $M$, in the potential field of the
$SU(2)$-soliton. The bound state solutions to this wave equation
represent stable hyperon states. \\

In the harmonic approximation the interactions between the mesons in
the hyperons, that are formed as multimeson states as the cascade and
$\Omega^{-}$ particles, are neglected. We here carry out the expansion
in $M$ to quartic order in order to obtain the lowest anharmonic
correction to the Hamiltonian density for the meson field $M$. This
takes the form of an effective meson-meson interaction, which we treat
as a first order perturbation. The employment of the alternate quartic
stabilizing term (2.2) is motivated by the fact that it leads to a
much simpler form for the anharmonic correction than Skyrme's original
stabilizing term. We show that the alternate term leads to
binding energies for the heavy isodoublet pseudoscalar mesons,
which are very close to those obtained with
the Skyrme term, and hence it should suffice for the purpose of
estimating the magnitude of the anharmonic correction. If a realistic
description of the interaction between the soliton and the complete
pseudoscalar nonet is required one has in general to consider a linear
combination of the two quartic stabilizing terms [13].\\

In the expansion of the Lagrangian density (2.1) in powers of the meson
field $M$ the lowest (zero) order term is the usual Lagrangian density
of the $SU(2)$ soliton [16]. Because of the stability of the soliton
only terms of even order in $M$ appear. To second order one obtains in
the case of an unrotated soliton field [1,7] the following
expressions:

$${\cal L}_\sigma^{(2)}={1\over \chi^2}\partial_\mu M^\dagger \partial^\mu M
+{1\over \chi^2}M^\dagger
\{(2\frac{sin^2\theta/2}{r^2}\cos\theta+{\theta^{'2} \over 4})$$
$$-\frac{sin^2\theta/2}{r^2}\vec \tau \cdot \vec L\}M,\eqno(2.11a)$$

$${\cal L}_S^{(2)}={1\over e^2f^2}\{(\theta^{'2}+{sin^2\theta \over
r^2})\partial_\mu M^\dagger \partial^\mu M$$
$$+(\theta^{'2}-{sin^2\theta \over r^2})\hat r \cdot \nabla M^\dagger
\quad \hat r \cdot \nabla M$$
$$+M^\dagger[2cos \theta {sin^2\theta/2 \over
r^2}(\theta^{'2}+{sin^2\theta \over r^2})+{\theta^{'2} \over
2}sin^2\theta$$
$$-2{sin^2\theta/2 \over r^2}(\theta^{'2}+{sin^2\theta \over r^2})\vec
\tau \cdot \vec L]M\},\eqno(2.11b)$$

$${\cal L}_{CSB}^{(2)}=-m^2M^\dagger M
+(\chi^2-1){\cal L}_\sigma^{(2)}
.\eqno(2.11c)$$
Here the factor $\chi$ is the ratio $f/f_\pi$.
Finally the Wess-Zumino term (2.6) gives rise to the following
contribution to the meson Lagrangian in the harmonic approximation
[2]:

$${\cal L}_{WZ}=i{N_c\over 4f^2} B^\mu(M^\dagger D_\mu M-(D_\mu
M)^\dagger M).\eqno(2.12)$$
Here $B^\mu$ is the anomalous baryon current of the soliton [16], the
time component of which has the explicit form

$$B^0=-{1\over 2\pi^2r^2}sin^2\theta \, \theta',\eqno(2.13)$$
and $D_\mu$ is the "covariant" derivative [1]

$$D_\mu\equiv \partial_\mu+{1\over 2}(\sqrt{U_\pi^\dagger}\partial_\mu
\sqrt{U_\pi} + \sqrt{U_\pi} \partial_\mu
\sqrt{U_\pi^\dagger}).\eqno(2.14)$$
The second order Lagrangian densities (2.11), (2.12) do, upon a mode
decomposition of the kaon field, lead to the following wave equation
for the bound meson:

$$a(r)\nabla^2k+b(r)\hat r \cdot \nabla k-c(r)\vec L^2k$$
$$-[v_0(r)+v_{IL}(r)\vec I \cdot \vec L]k-m^2k+2\omega \lambda (r)k$$
$$+d(r)\omega^2 k=0.\eqno(2.15)$$
Here $\omega$ is the energy of the bound meson $(\omega < m)$
and $k(\vec r)$ is the meson wave function. The
effective (iso)spin operator $\vec I$ for the system is defined as
$\vec I=\vec \tau/2$. The radial functions $a,...\lambda$ in eqn.
(2.15) are defined as

$$a(r)=1+{2\over e^2f^2}{sin^2\theta \over r^2},\eqno(2.16a)$$
$$b(r)=-{4\over e^2f^2}{sin\theta \over r^2}({sin\theta \over r}-cos
\theta\, \theta'),\eqno(2.16b)$$
$$c(r)={1\over e^2f^2r^2} (\theta^{'2}-{sin^2\theta \over
r^2}),\eqno(2.16c)$$
$$d(r)=1+{1\over e^2f^2}(\theta^{'2}+2{sin^2 \theta \over
r^2}),\eqno(2.16d)$$
$$v_0(r)=-2cos\theta{sin^2\theta/2 \over r^2}-{\theta^{'2} \over 4}-{1\over
2e^2f^2}\{\theta^{'2} {sin^2\theta \over r^2}$$
$$+4\frac{sin^2\theta/2\,
cos\theta}{r^2}(\theta^{'2} +{sin^2\theta \over r^2})\},\eqno(2.16e)$$
$$v_{IL}(r)=4{sin^2\theta/2 \over r^2}+{4\over e^2f^2}{sin^2\theta/2
\over r^2}(\theta^{'2}+{sin^2\theta \over r^2}), \eqno(2.16f)$$
$$\lambda(r)=-\frac{N_c}{8\pi^2f^2} \frac{sin^2\theta \,
\theta'}{r^2}.\eqno(2.16g)$$

The ground state solution to the meson wave equation (2.15) is a
$P$-state. In the case of the kaon we obtain the ground state energy
to be 195 MeV when $f_K=1.23 f_\pi$, and the parameter values $f_\pi$
and $e$ are chosen as in ref. [16] so that the nucleon and
$\Delta_{33}$ resonance have their empirical mass values $(f_\pi=64.5$
MeV, $e$=5.45). This value for the energy of the bound $K$ meson leads
to satisfactory predictions for the masses of the stable strange
hyperons [4].\\

In the case of the heavier flavour hyperons the alternate quartic
stabilizing term (2.11b) does not lead to as satisfactory predictions
for the energies of the charm and bottom hyperons if the decay
constant ratios $f_D/f_\pi$ and $f_B/f_\pi$ are varied within the
presently accepted uncertainty range. Unless these decay constant
ratios are given fairly large values the $D$ and $B$ mesons will be
predicted to be overbound by a significant amount. We shall here use
the value $f_D/f_\pi=2.4$, which gives the value $\omega_D=1226$ MeV,
which is about 100 MeV too small. If the value for $f_D/f_\pi$ is
reduced to the value 1.8 at the upper end of the empirical uncertainty
range the predicted value for $\omega_D$ drops to only 996
MeV, which is unrealistically low. In the case of the $B$ meson we
obtain $\omega_B$=3528 MeV with the value $f_B/f_\pi$=2.8, which is
somewhat above the presently accepted uncertainty range. This value
for $\omega_B$ is about 1 GeV below the more realistic value that is
obtained with the original Skyrme model stabilizing term [4], which
reinforces the conclusion that the alternate quartic stabilizing
term should not be used
for phenomenological predictions except in combination with Skyrme's
stabilizing term [13]. We
nevertheless assume that it is sufficiently realistic to permit an
estimate of the relative importance of the anharmonic corrections to
the energy.

\vspace{1cm}

{\bf 2.2 The anharmonic Hamiltonian density}
\vspace{0.5cm}

We now turn to the derivation of the lowest order anharmonic
correction to the Lagrangian density (2.11), (2.12) for the meson
field by expanding the full Lagrangian density (2.1) to quartic order
in the kaon fields. The quartic anharmonic correction that arises from
the non-linear $\sigma$-model term (2.2) has the following form:

$${\cal L}_\sigma^{(4)}={1\over 6\chi^2f^2}\{\partial_\mu M^\dagger
M\partial^\mu M^\dagger M+M^\dagger \partial_\mu M M^\dagger
\partial^\mu M$$
$$-M^\dagger M\partial_\mu M^\dagger \partial^\mu M-M^\dagger
\partial_\mu M\partial^\mu M^\dagger M\}$$

$$-{1\over 12 \chi^2f^2}\{6M^\dagger \vec l M\cdot M^\dagger \vec
rM+M^\dagger (\vec l\cdot \vec r+\vec r \cdot \vec l)MM^\dagger M$$
$$+M^\dagger M[M^\dagger(\vec l+\vec r) \cdot \vec \nabla M-\vec
\nabla M^\dagger  \cdot (\vec l+\vec r) M]$$
$$+3M^\dagger (\vec l+\vec r)M\cdot [M^\dagger \vec \nabla M-\vec
\nabla M^\dagger M]\}.\eqno(2.17)$$
Here $\vec l$ and $\vec r$ are defined as the left and right currents
for the square root of the soliton field:

$$\vec l=\sqrt{U_\pi^\dagger}\vec \nabla \sqrt{U_\pi},\eqno(2.18a)$$
$$\vec r=\sqrt{U_\pi}\vec \nabla \sqrt{U_\pi^\dagger}.\eqno(2.18b)$$
The explicit forms of these are

$$\vec l=i\{{sin\theta \over 2r}\vec \tau +{sin^2\theta/2 \over r}
\vec \tau \times \hat r +({\theta' \over 2}-{sin \theta \over 2r})\vec
\tau \cdot \hat r\hat r\},\eqno(2.19a)$$
$$\vec r=-i\{{sin\theta \over 2r} \vec \tau- {sin^2\theta/2 \over r}
\vec \tau \times \hat r+({\theta' \over 2}-{sin \theta \over 2r})\vec
\tau \cdot \hat r\hat r\}.\eqno(2.19b)$$

The quartic anharmonic correction that arises from the stabilizing
term ${\cal L}_S$ (2.4) can be expressed in the following form:

$${\cal L}_S^{(4)}={1\over 4e^2 f^4}\{(\theta^{'2}+{sin^2\theta
\over r^2})A_{\mu\mu}-(\theta^{'2}-{sin^2\theta \over r^2})\hat r_m\hat
r_nA_{mn}\}$$
$$+{1\over 16e^2f^4}\{B_{\mu \nu}B^{\mu
\nu}-(B_{\mu\mu})^2\}.\eqno(2.20)$$
Here we have introduced two tensors $A_{\mu\nu}$ and $B_{\mu\nu}$ which are
defined as follows:

$$A_{\mu\nu}={1\over 3}\{M^\dagger M(\partial_\mu M^\dagger \partial_\nu
M+\partial_\nu M^\dagger \partial_\mu M)$$
$$+\partial_\nu M^\dagger M M^\dagger \partial_\mu M
+\partial_\mu M^\dagger M
M^\dagger \partial_\nu M$$
$$-M^\dagger \partial_\nu M M^\dagger \partial_\mu M-M^\dagger
\partial_\mu M
M^\dagger \partial_\nu M$$
$$-\partial_\mu M^\dagger M \partial_\nu M^\dagger M-\partial_\nu M^\dagger
M\partial_\mu M^\dagger M$$
$$-2M^\dagger [\frac{cos\theta sin^2\theta/2}{r^2}\delta_{mn}+({\theta^{'2}
\over 4}-\frac{cos \theta sin^2\theta/2 }{r^2})\hat r_m \hat
r_n]MM^\dagger M$$
$$-6M^\dagger[{sin\theta \over 2r}\tau_m +({\theta^{'} \over 2}
-{sin\theta \over 2r})\vec \tau \cdot \hat r \hat r_m] MM^\dagger[{sin \theta
\over 2r} \tau_n+({\theta'\over 2}-{sin\theta \over 2r})\vec \tau
\cdot \hat r \hat r_n]M$$
$$+6M^\dagger \kappa(r)(\vec \tau \times \hat r)_mMM^\dagger
\kappa(r)(\vec \tau \times \hat r)_nM$$
$$-iM^\dagger \kappa(r)(\vec \tau \times \hat r)_m \nabla_n M
M^\dagger M$$
$$+i\nabla_n M^\dagger \kappa(r)(\vec \tau \times \hat r)_m MM^\dagger
M$$
$$-iM^\dagger \kappa(r) (\vec \tau \times \hat r)_n \nabla_mM M^\dagger
M$$
$$+i\nabla_m M^\dagger \kappa(r) (\vec \tau \times \hat r)_nMM^\dagger
M$$
$$+3i M^\dagger \kappa(r)(\vec \tau \times \hat
r)_mM(\nabla_nM^\dagger M-M^\dagger \nabla_n M)$$
$$+3M^\dagger \kappa(r)(\vec \tau \times \hat r)_nM(\nabla_m
M^\dagger M-M^\dagger \nabla_mM)\},\eqno(2.21)$$
and

$$B_{\mu \nu}=-2(\partial_\mu M^\dagger \partial_\nu M+\partial_\nu
M^\dagger \partial_\mu M)$$
$$+M^\dagger (l_m r_n+r_ml_n+l_nr_m +r_ml_n)M$$
$$+2i \kappa(r)[M^\dagger (\tau \times \hat r)_m\partial_\nu M+M^\dagger
(\vec \tau \times \hat r)_n\partial_\mu M$$
$$-\partial_\mu M^\dagger (\vec \tau \times \hat r)_nM-\partial_\mu M^\dagger
(\vec \tau \times \hat r)_mM].\eqno(2.22)$$
The indices $m$ and $n$ in the expression for $B_{\mu\nu}$ indicate the
spatial components of $B$. Above we have used the notation
$$\kappa(r)={sin^2\theta/2\over r}.\eqno(2.23)$$

Finally the charge symmetry breaking Lagrangian (2.5) gives rise to
the following quartic anharmonic term:

$${\cal L}_{CSB}^{(4)}={2\over 3} {m^2\over f^2} M^\dagger
MM^\dagger M$$
$$ -{1\over 2f^2}(1-{1\over \chi^2})\{{4\over 3} K^\dagger
K\partial_\mu K^\dagger \partial^\mu K+{4\over 3}\partial_\mu
K^\dagger K K^\dagger \partial^\mu K$$
$$-{1\over 3}K^\dagger \partial_\mu K K^\dagger \partial^\mu K-{1\over
3}\partial_\mu K^\dagger K \partial^\mu K^\dagger K$$
$$+{2\over 3}K^\dagger KK^\dagger(l_m
r_m+r_ml_m)K$$
$$-{2\over 3} K^\dagger K[(\nabla_m
K^\dagger(l_m+r_m)K-K^\dagger(l_m+r_m)\nabla_m K)]$$
$$+(K^\dagger \nabla_m K-\nabla_m K^\dagger K)K^\dagger(l_m+r_m)K$$
$$+2K^\dagger l_m KK^\dagger r_m K\} .\eqno(2.24)$$

The quartic anharmonic Hamiltonian density is obtained by introduction
of conjugate momentum fields as
$$\Pi=\frac{\partial{\cal L}}{\partial \dot M^\dagger},\quad
\Pi^\dagger=\frac{\partial{\cal L}}{\partial \dot M}, \eqno(2.25)$$
and carrying out the Legendre transformation
$${\cal H}=\Pi^\dagger\dot M+\dot M^\dagger\Pi-{\cal L}.\eqno(2.26)$$
The resulting expression for the anharmonic term in ${\cal H}$
is identical to that of the quartic lagrangian densities (2.17),
(2.20) and (2.24), with exception for an overall sign change
of those terms that do not contain time derivatives of the meson
field or its Hermitean conjugate.

The quartic Hamiltonian density has the form of meson-meson interaction.
In general the interacting mesons can be in different states,
although we shall here consider the case of both mesons in
the ground state. As the ground state is a P-state it is described
by a wave function of the form
$$M=A\frac{k(r)}{\sqrt{4\pi}}\vec\tau\cdot\hat r,\eqno(2.27)$$
where $k(r)$ denotes the radial wave function and $A$ is the rotation
operator, which transforms the meson isospin operator into an effective
spin operator. As the soliton field is rotated by the same $SU(2)$
rotation operator, i.e. $u\rightarrow AuA^\dagger$, the isospin
operator $\vec\tau$ in the Lagrangian densities is converted into
the spin operator of the meson as well. Note that the normalization
condition for the meson wave function $k(r)$ is
$$2\int_0^\infty drr^2k^2(r)[\omega d(r)+\lambda(r)]=1,\eqno(2.28)$$
a result that is implied by the form of the harmonic Lagrangian
for the meson.\\

The anharmonic correction to the energy of a two-meson state, which
corresponds to a hyperon of the cascade type, is obtained as
$$\Delta E^{(4)}=\int d^3r{\cal H}^{4},\eqno(2.29)$$
where ${\cal H}^{4}$ is the Hamiltonian density of quartic order in $M$.
The expression for this Hamiltonian density has  the general form

$${\cal H}^{(4)}=M^\dagger O_1MM^\dagger O_2 M,\eqno(2.30)$$
where $O_j$ are operators that depend on the chiral angle $\theta$
of the soliton and the spin and radial coordinates of the mesons.
In the matrix element (2.29) the meson fields M are replaced by
the corresponding wave functions, which in the case of both mesons
being in the ground state have the form (2.27).
The general expression for the anharmonic correction (2.29)
will then be
$$\Delta E^{(4)}=\frac{2}{(4\pi)^2}\int d^3 r[k(r)\vec\sigma_1\cdot
\hat r O_1\vec\sigma_1\cdot\hat rk(r)][k(r)\vec\sigma_2\cdot\hat r
O_2 \vec\sigma_2 \cdot\hat r k(r)].\eqno(2.31)$$
Note that in the
case when both mesons are in the same orbital symmetry they form
a spin triplet state because they satisfy Bose statistics.\\

\vspace{1cm}

\centerline{\bf 3. The anharmonic correction for cascade hyperons}
\vspace{0.5cm}

In the case of two-meson states, when both mesons have the same
flavour and are in the ground state (2.27) the expression for the
matrix element of the anharmonic Hamiltonian derived above can be
reduced to a very compact form, as in this case $\vec \tau \cdot \vec
L=-2$ and $\vec \sigma^1 \cdot \vec \sigma^2=1$ as the mesons form a
triplet state. The anharmonic correction that arises from the
quadratic term in the Hamiltonian, which is obtained by converting
${\cal L}_\sigma^{(4)}$ (2.17) into a corresponding Hamiltonian
density is

$$\Delta E_\sigma^{(4)}=-{1\over 3\pi \chi^2 f^2}\int dr\,
r^2k^4(r)\{\omega^2+{1\over r^2}g(r)-{\theta^{'2} \over 8}\},\eqno(3.1)$$
where the function $g(r)$ is defined as

$$g(r)=cos \theta cos^2{\theta \over 2}.\eqno(3.2)$$

The correction to the energy that arises from the quartic stabilizing
term ${\cal L}_s^{(4)}$ (2.20) takes the following form:

$$\Delta E_S^{(4)}={1\over 3\pi e^2f^4} \int dr \, k^2(r)$$
$$\{k^2(r)[{g(r) \over r^2}(3g(r)-sin^2 \theta)+{\theta^{'2} \over
4}(2 g(r) +sin^2\theta)]$$
$$-6k^{'2}(r)g(r)\}$$
$$-{\omega^2\over
2}[k^2(r)(3sin^2\theta+
\frac{7}{2}r^2\theta^{'2}+4cos^2 {\theta\over 2} -12
g(r))$$
$$-6r^2 k^{'2}(r)]\}.\eqno(3.3)$$

Finally the chiral symmetry breaking term
${\cal L}_{CSB}^{4}$ (2.24) yields the correction

$$\Delta E_{CSB}^{(4)}=-\frac{1}{3\pi f^2}\int dr r^2 k^2(r)\{
m^2 k^2(r) $$
$$-(1-\frac{1}{\chi^2})[k^2(r)[-{5\over 2}\omega^2+{3\over 8} \theta^{'2}+
-{3\over r^2}g(r)]$$
$$-{3\over 2} k^{'2}]\}. \eqno(3.4)$$

Note that all the coefficients of the wave functions in these
expressions are nonsingular at the origin.\\

Numerically we find the anharmonic quartic correction to the energy of
a two-kaon state to be -26 MeV (Table I). This represents only a 7\% increase
of the energy 390 MeV that is obtained in the harmonic approximation
for the two meson state.
The relative smallness of the anharmonic correction indicates that the
harmonic approximation is sufficiently reliable to use in the study of
hyperon structure and that thus the fairly satisfactory prediction for
the spectra [4] and magnetic moments [18-20] of the strange hyperons
have uncertainties of no more than 10\% due to the anharmonic terms.
Note that the 10\% correction in the meson energy corresponds to only
a 2\% correction to the predicted energies of the cascade particles,
and therefore may be viewed as insignificant. The contributions
$\Delta E_\sigma^{(4)}$, $\Delta E_S^{(4)}$ and $\Delta E_{CSB}^{(4)}$
are listed separately in Table I. Of the value -32 MeV for $\Delta
E_{CSB}^{(4)}$ the mass term alone contributes -29 MeV, and it thus
represents the dominant term.\\

The anharmonic correction to the energy of a two $D$-meson state is
-233 MeV (Table I). As the energy of the bound two $D$-meson state is
2452 MeV this
represents a correction of only 10 percent. In the case of the $B$-meson we
find
with $\chi=2.8$ the bound state energy 7056 MeV. In this case the
anharmonic correction to be -639 MeV and transfer less significant than in
the case of the charmed cascade hyperon. The separate contributions to
the total anharmonic correction are given in Table I.\\

The fact that the relative magnitude of the anharmonic correction does
not increase with
increasing meson mass may be understood with the help of the following
argument. At large distances the meson wave function behaves as

$$k(r)=N\frac{e^{-\sqrt{m^2-\omega^2}r}}{r},\eqno(3.5)$$
where $N$ is a constant.
In the case of the ground state (2.27) $k(r)$ is finite and
nonzero at the origin.
The following approximate form for the meson wave function

$$k(r)\simeq N e^{-\sqrt{m^2-\omega^2}r},\eqno(3.6)$$
has the correct behaviour at the origin and exponential tail at large
distances. The normalization constant $N$ is determined by the
normalization condition [1]

$$2\int dr r^2k^2(r)(\omega d(r)+\lambda (r))=1.$$
For the wave function model (3.6) this implies that

$$N \sim \frac{(m^2-\omega^2)^{3/4}}{\sqrt{\omega}}\eqno(3.7)$$
when the meson mass and energy is large $(\omega >> \lambda (0))$. For
large energies the main anharmonic contribution is obtained from those
terms in (3.1), (3.3) and (3.4) which are proportional to $m^2$ (the
mass term) and $\omega^2$.
If the soliton functions these terms in the integrands are replaced by
constants they yield the estimate

$$\Delta E^{(4)}\sim {N^4\over
f^4}\frac{\omega^2}{(m^2-\omega^2)^{3/2}}.\eqno(3.8)$$
Here we have used the fact that $\omega \sim m$ for large meson mass values.
By (3.7) it then follows that

$$\Delta E^{(4)} \sim \frac{(m^2-\omega^2)^{3/2}}{f^4}.\eqno(3.9)$$
In ref. [14] it has been shown that the consistency of the soliton
model requires that

$$f\sim \sqrt{m},\eqno(3.10)$$
and in ref. [21] that to a good approximation for large meson mass
values the meson energy is given as

$$\omega=\frac{m}{\sqrt{1+{3\over \chi^2}}}.\eqno(3.11)$$
Combination of these results yield the following scaling law for the
anharmonic correction:

$$\Delta E\sim {1\over \sqrt{m}},\eqno(3.12)$$
which shows that the anharmonic correction falls with meson mass and
would become insignificant in the case of very large meson mass
values.\\

\vspace{1cm}

{\bf 4. Discussion}
\vspace{0.5cm}

The results presented above for the anharmonic correction to the energy of a
bound two-meson state show that it reduces the predicted energy
in the harmonic approximation for a two meson state
by 10 \%. This indicates that the
harmonic approximation is reliable for the strange, charm and bottom
hyperon generations, and
that the satisfactory results it leads to for the static properties
of the heavy flavour hyperons are fairly stable. As the topological model
implies a large colour number approximation its inherent accuracy
should in any case not be expected to be better than about 10\%. It
should also be noted that small adjustments of the values of the
decay constants that appear in the Lagrangian density can lead to
shifts in the predicted meson energies that are much larger than
10\% [3].\\

The numerical results in this work apply only to the model with the
alternate quartic stabilizing term (2.4). But the qualitative conclusion
that the anharmonic corrections to the hyperon energies are small is not
restricted to this particular version of the soliton model. It has been
shown [14] that models with alternate stabilizing terms give qualitatively
similar results in the
harmonic approximation. The consistency argument for the smallness of
the anharmonic
correction given in section 3 is also general.\\

The anharmonic terms do also in principle contribute to the energy of
a single meson state in the form of self energy corrections. These
corrections, which have not been considered here, require summation
over all meson states and hence a method of regularizing the high
energy terms in the sum. As the Lagrangian model (2.1) is constructed
as an effective low energy interaction the evaluation of these self
energy corrections lead outside the range of its validity and would
require a systematic approach along the lines of chiral perturbation
theory.

\newpage

{\bf References}\\
\vspace{0.5cm}

\begin{enumerate}
\item C.G. Callan and I. Klebanov, Nucl. Phys. {\bf B262} (1985) 365
\item C. Callan, K. Hornbostel and I. Klebanov, Phys. Lett. {\bf B202}
(1988) 269
\item M. Rho, D.O. Riska and N.N. Scoccola, Phys. Lett. {\bf B251}
(1990) 597
\item M. Rho, D.O. Riska and N.N. Scoccola, Z. Phys. A {\bf 341}
(1992) 343
\item N.N. Scoccola, D.P. Min, H. Nadeau and M. Rho, Nucl. Phys. {\bf
A505} (1989) 497
\item E. Jenkins, A.V. Manohar and M.B. Wise, Nucl. Phys. {\bf B396}
(1993) 27
\item C.L. Schat and N.N. Scoccola, Preprint TAN-FNT-93-17 (1993)
\item Y. Oh, B.Y. Park and D.P. Min, Preprint SNUT-93/80 (1993)
\item D.O. Riska and N.N. Scoccola, Phys. Lett. {\bf B299} (1993) 338
\item T.H.R. Skyrme, Proc. Roy. Soc. {\bf A260} (1961) 127
\item U. Blom K. Dannbom and D.O. Riska, Nucl. Phys. {\bf A493} (1989)
384
\item G. Pari, Phys. Lett. {\bf B261} (1991) 347
\item G. Pari and N.N. Scoccola, Phys. Lett. {\bf B296} (1992) 391
\item M. Bj\"{o}rnberg, K. Dannbom, D.O. Riska and N.N. Scoccola, Nucl.
Phys. {\bf A539} (1992) 662
\item G. Pari, B. Schwesinger and H. Walliser, Phys. Lett. {\bf B255}
(1991) 1
\item G.S. Adkins, C.R. Nappi and E. Witten, Nucl. Phys. {\bf B228}
(1983) 552
\item D.O. Riska and N.N. Scoccola, Phys. Lett. {\bf B265} (1991) 188
\item E.M. Nyman and D.O. Riska, Nucl. Phys. {\bf B325} (1989) 593
\item J. Kunz and P.J. Mulders, Phys. Rev. {\bf D41} (1990) 1578
\item Y. Oh, D.-P. Min, M. Rho and N.N. Scoccola, Nucl. Phys. {\bf
A534} (1991) 493
\item M. Bj\"{o}rnberg and D.O. Riska, Nucl. Phys. {\bf A549} (1992) 537
\end{enumerate}

\newpage

\centerline{\bf Table I}
\vspace{0.4cm}

Numerical values (in MeV) for the different anharmonic contributions
to the energy of a 2-meson state with the mesons in the ground state.

\vspace{1cm}
\begin{center}

\begin{tabular}{|c|c|c|c|}\hline
 & K & D & B\\ \hline
$\Delta E_\sigma^{(4)}$ & 3 & --6 & --21\\
$\Delta E_S^{(4)}$ & 3 & --44 & 144\\
$\Delta E_{CSB}^{(4)}$ & --32 & --183 & --762\\ \hline
$\Delta E_{TOT}^{(4)}$ & --26 & --233 & --639\\ \hline
\end{tabular}
\end{center}

\end{document}